\documentclass[useAMS,usenatbib]{mnras}
\usepackage[export]{adjustbox}
\usepackage{amssymb,amsmath}
\usepackage{color}
\usepackage{epstopdf}
\usepackage{natbib}
\usepackage{verbatim}
\usepackage{lipsum}
\usepackage[T1]{fontenc}

\title[Environment effects on the shape of galaxies]{The effects of environment on the intrinsic shape of galaxies}
\author[S. Rodr\'iguez, N.D. Padilla and D. Garc\'ia Lambas]{Silvio Rodr\'iguez$^1$\thanks{E-mail:sirodrig@uc.cl}, Nelson D. Padilla$^{2,3}$ and Diego Garc\'ia Lambas$^{1,4}$\\
$^{1}$Instituto de Astronom\'ia Te\'orica y Experimental, UNC-CONICET, C\'ordoba, Argentina\\
$^{2}$Instituto de Astrof\'isica, Pontificia Universidad Cat\'olica de Chile, Santiago, Chile\\
$^{3}$Centro de Astro-Ingenier\'ia, Pontificia Universidad Cat\'olica de Chile, Santiago, Chile\\
$^{4}$Observatorio Astron\'omico de C\'ordoba, Universidad Nacional de C\'ordoba, Argentina}

\begin{document}

\date{Accepted 2015 November 10.  Received 2015 November 4; in original form 2015 April 21}

\pagerange{\pageref{firstpage}--\pageref{lastpage}} \pubyear{2015} %cuidado aqui

\maketitle

\label{firstpage}

\begin{abstract}
    We measure the effect of the environment on the intrinsic shapes of spiral and elliptical galaxies by finding the 3D shape distribution and dust extinction that fits better the  projected shape of galaxies in different environment. We find that spiral galaxies in groups are very similar to field spirals with similar intrinsic properties (magnitudes, sizes and colours). But for spirals in groups, those in denser environments or closer to the centre of the group tend to have a more circular disc than similar galaxies in less dense environments or far from the group centres. Also we find that central spiral galaxies in their groups tend to be thinner than other similar spirals.
    
	For ellipticals, we do not find any important dependence of their shape on their position in a group or on the local density. However, we find that elliptical galaxies in groups tend to be more spherical than field ellipticals with similar intrinsic properties.
 
	We find that, once in groups, the shape of member galaxies do not depend on group mass, regardless of their morphological type.

\end{abstract}

\begin{keywords}
  galaxies: structure -- galaxies: general -- galaxies: fundamental parameters -- galaxies: groups: general -- surveys
\end{keywords}

\section{Introduction}
\label{intro}

It is well known that the environment has observable effects on the galaxies, either on the star formation history \citep{sfh_env0,sfh_env}, gas content \citep{gas_env0,gas_env}, morphology \citep{mor_env,dressler}, structure \citep{str_env, str_env2}, etc. The location of a galaxy regarding large scale structure plays an undeniable role in his development.

The study of galaxy shapes can shed light about their dynamics \citep{davies,binney}, and also bring information on their formation history \citep{sandage}. Also, the knowledge of the realistic distribution of galaxy shapes can help to test both semi-analytic and hydrodynamical simulated galaxy formation models. Besides, the agreement between the shape distribution of simulated and real galaxies serves as a further test of the models. In particular, it is important to study the dependence of the shape of galaxies on environment if we aim at fully understanding the processes that drive the formation and dynamics of the stellar content in galaxies.

The effects of the environment on the morphology of a galaxy were first pointed out by \cite{dressler}. He found that there is a clear relation between the local density and the morphological type. As the local density increases, the abundance of spirals decreases and that of ellipticals and S0 increases. He also found that for ellipticals the luminosity tends to increase with the local density.

More recently, \cite{kuehn} concentrated on the possible impact of the environment on the shape of galaxies. They found that galaxies with exponential luminosity profiles tend to be flatter in high density environments and that galaxies with a de Vaucouleurs luminosity profile with similar magnitudes tend to be rounder in high density environments, but \citeauthor{kuehn} only used the projected shape ($b/a$) to perform their analysis, and they did not restrict their samples to the same intrinsic properties.

The intrinsic shape of galaxies has been analysed in many works \citep[][etc]{sandage,lambas,andersen}, one of the most recent is \citeauthor{rp13} (2013, hereafter RP13), a work that extended the analysis made by \cite{nym} using data from Sloan Digital Sky Survey (SDSS) Data Release 8 \citep{dr8} and Galaxy Zoo 1\citep{lintott2}.  RP13 take into account the parametrizations of galaxy shapes in \cite{ssg}.

The goal of this work is to extend the work in RP13 with environmental information, that is, to determine if the distribution of intrinsic shapes of the galaxies depends on the environment of the galaxies. A priori, it would be expected to find galaxies with more spherical shapes in denser environments, specially in the case of ellipticals. For spirals we could expect a similar effect although late types are significantly less frequent in denser, environments so our data could not exhibit a significant trend.

RP13 models the distribution of $\gamma \equiv C/A$ and $\epsilon \equiv 1-B/A$, where $A$, $B$ and $C$ are the intrinsic minor, medium and major axis of a galaxy, respectively, for samples of spirals, ellipticals and sub-samples selected by magnitude, size and colour. They use the $\gamma$ and $\epsilon$ distribution, along with the dimming of the magnitude in edge-on galaxies due to dust ($E_0$) to fit the observed $b/a$ distribution, where $b$ and $a$ are the observed minor and major axis of a galaxy. The relation between $\gamma$ and $\epsilon$ and $b/a$ is the same given in \cite{binney}.

$E_0$ is used to model the distribution of the polar inclination angle $\theta$ in the sample of spirals (in ellipticals they assume $E_0=0$, that is, a flat $\cos \theta$ distribution), due to the fact that, in a flux limited sample, the number of edge-on galaxies detected is smaller than the number of face-on galaxies due to the dust obscuration. The value of $E_0$ is obtained from the difference in the luminosity function of the edge-on and face-on samples. The galaxies that belong to each sample are chosen according to the $b/a$ value.

To model $\gamma$ and $\epsilon$, RP13 use two different types of parametrization. In the case of ``Type n" they parametrize the $\gamma$ distribution as the sum of 10 Gaussians, with dispersion 0.08 and centres running from 0.1 to 0.82, with steps of 0.08, varying the number of galaxies in each Gaussian to change the shape of the distribution. The $\epsilon$ distribution is parametrized as the sum of the positive side of 10 Gaussians \citep[each one weighted by its own dispersions, see][]{errata} centred in 0, with dispersions running from 0.02 to 0.2 with steps of 0.02. In the case of ``Type r" they also parametrize $\gamma$ as the sum of 10 Gaussians with dispersion of 0.08, but the centres run from 0.04 to 0.4 with steps of 0.04. The $\epsilon$ distribution is parametrized as a log-normal distribution, with variable centre and dispersion. The sub-samples of ellipticals are well fitted with the parametrization type n; in the case of spirals it depends on each sub-sample. In particular, the total sample of spirals is well fitted with type r. This is the parametrization we will adopt in this letter.

This work is organized as follows. In \S \ref{sample} we describe the sample used in this work and how we select the sub-samples with different environment. In \S \ref{results} we show the main results obtained for the shapes of these different sub-samples.  In \S \ref{conclusions} we summarize and discuss our main results.

\section{Sample}
\label{sample}

RP13 use a sample obtained from SDSS DR8 \citep{dr8} with morphological information from Galaxy Zoo \citep{lintott,lintott2} with cuts in magnitude and redshift. To obtain environmental information, we extend the group catalogue from \cite{zapata} to the SDSS DR8.  \cite{zapata} used a Friend-of-Friends algorithm applied to SDSS DR6\citep{dr6} to find galaxy groups and also determine their properties \citep[see also][]{padilla}.
We add this group information to the RP13 sample for this letter work. In the process we added a more stringent cut in redshift (including only galaxies out to $z<0.1$) to avoid incompleteness of the sample, due to the fact that above this redshift the group sample only includes the most massive ones. 

We study different options to produce subsamples and keep those that show different $b/a$ distributions\footnote{Unless it is explicitly indicated, we use $b/a$ values measured in the SDSS $r$-band.}. The final sub-samples are composed by galaxies in groups and in the field. Additionally, galaxies in groups are separated between centrals and satellites, where a galaxy is considered a central if it is the brightest galaxy in the group, and satellite otherwise\footnote{To determine the brightest galaxy in a group, we use the full sample of galaxies in groups, not only the galaxies with morphological classification from Galaxy Zoo.}, by $r/r_{vir}$, the distance from a galaxy to the centre of mass of the group in units of the virial radius, by $M/M_{\sun}$, the virial mass of the group, and by $\Sigma_5$, the projected density to the fifth nearest neighbour of the galaxy. We also construct a sample adding the field and satellite samples together.

\begin{table*}
  \begin{center}
    \caption{Medians of magnitudes, colours and sizes for samples of Spirals and Ellipticals separated by environmental properties.\label{table:med_cut}}
    \begin{tabular}{lcccccr}
	    & \multicolumn{3}{c}{Medians for spirals} & \multicolumn{3}{c}{Medians for ellipticals} \\
      Sample & $M_r-5 \log \textmd{h}$ & $g-r$ & $R_{50}/$kpc h & $M_r-5 \log \textmd{h}$ & $g-r$ & $R_{50}/$kpc h \\\hline
      group         	      & -20.606 & 0.718 & 3.899 & -20.543 & 0.973 & 2.243 \\
      field         	      & -20.602 & 0.711 & 3.894 & -20.54  & 0.97  & 2.299 \\\hline
      centrals      	      & -20.603 & 0.718 & 3.807 & -20.547 & 0.97  & 2.226 \\
      satellites    	      & -20.6   & 0.719 & 3.897 & -20.547 & 0.974 & 2.287 \\
      satellite+field	      & -20.602 & 0.714 & 3.892 & -20.54  & 0.97  & 2.299 \\\hline
      $\Sigma_5>\Sigma_{5,m}$ & -20.226 & 0.724 & 3.417 & -20.832 & 0.977 & 2.711 \\
      $\Sigma_{5,m}>\Sigma_5$ & -20.225 & 0.723 & 3.382 & -20.833 & 0.976 & 2.713 \\\hline
      $r_m>r/r_{vir}$         & -20.204 & 0.722 & 3.353 & -20.777 & 0.975 & 2.622 \\
      $r/r_{vir}>r_m$         & -20.203 & 0.724 & 3.378 & -20.787 & 0.976 & 2.605 \\\hline
      $M/M_{\sun}>M_m$        & -20.217 & 0.721 & 3.397 & -20.851 & 0.976 & 2.733 \\
      $M_m>M/M_{\sun}$        & -20.218 & 0.718 & 3.388 & -20.853 & 0.976 & 2.713 \\\hline
    \end{tabular}
  \end{center}
\end{table*}

\begin{figure*}
	\begin{center}
		\includegraphics[width=.215\textwidth,valign=t]{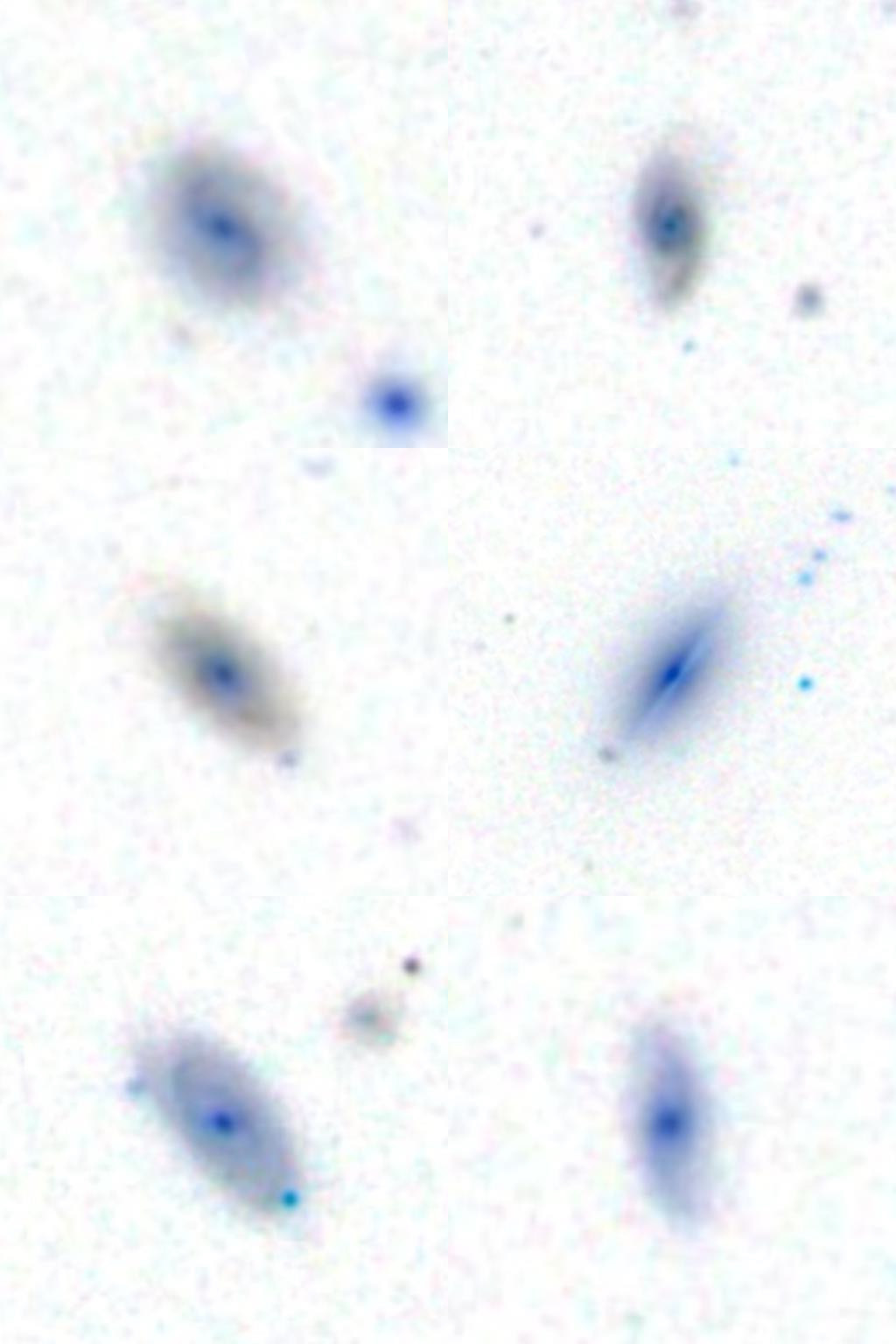}
		\includegraphics[width=.44\textwidth,valign=t]{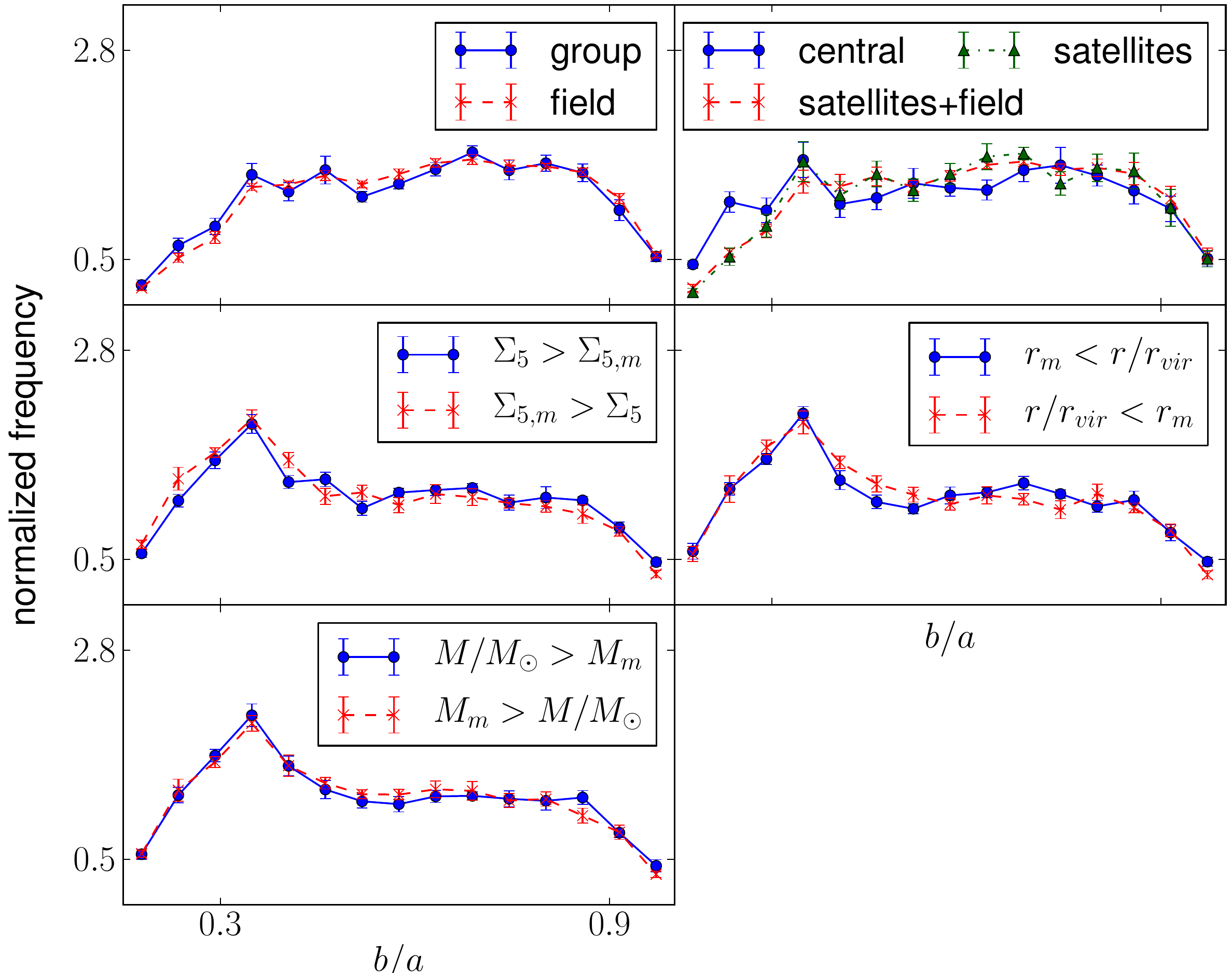}
		\includegraphics[width=.215\textwidth,valign=t]{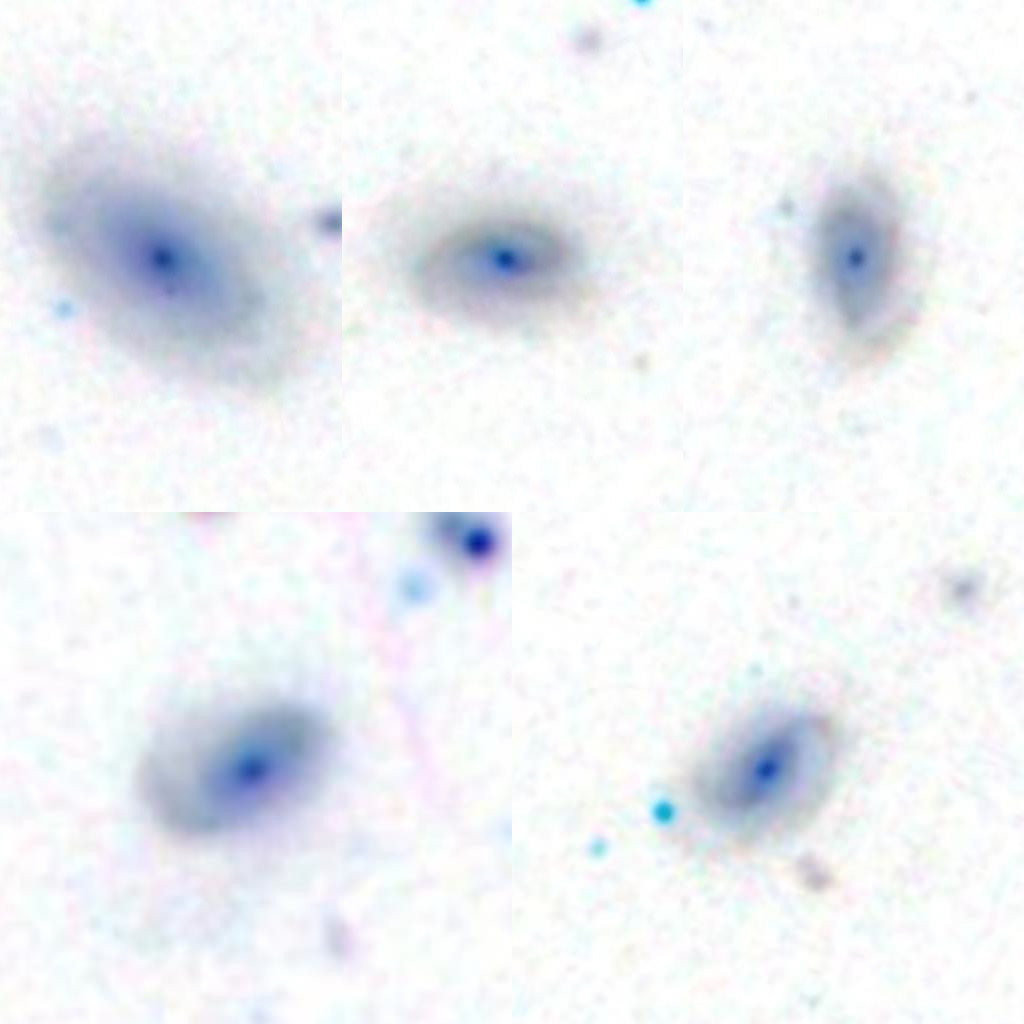}
		\includegraphics[width=.215\textwidth,valign=t]{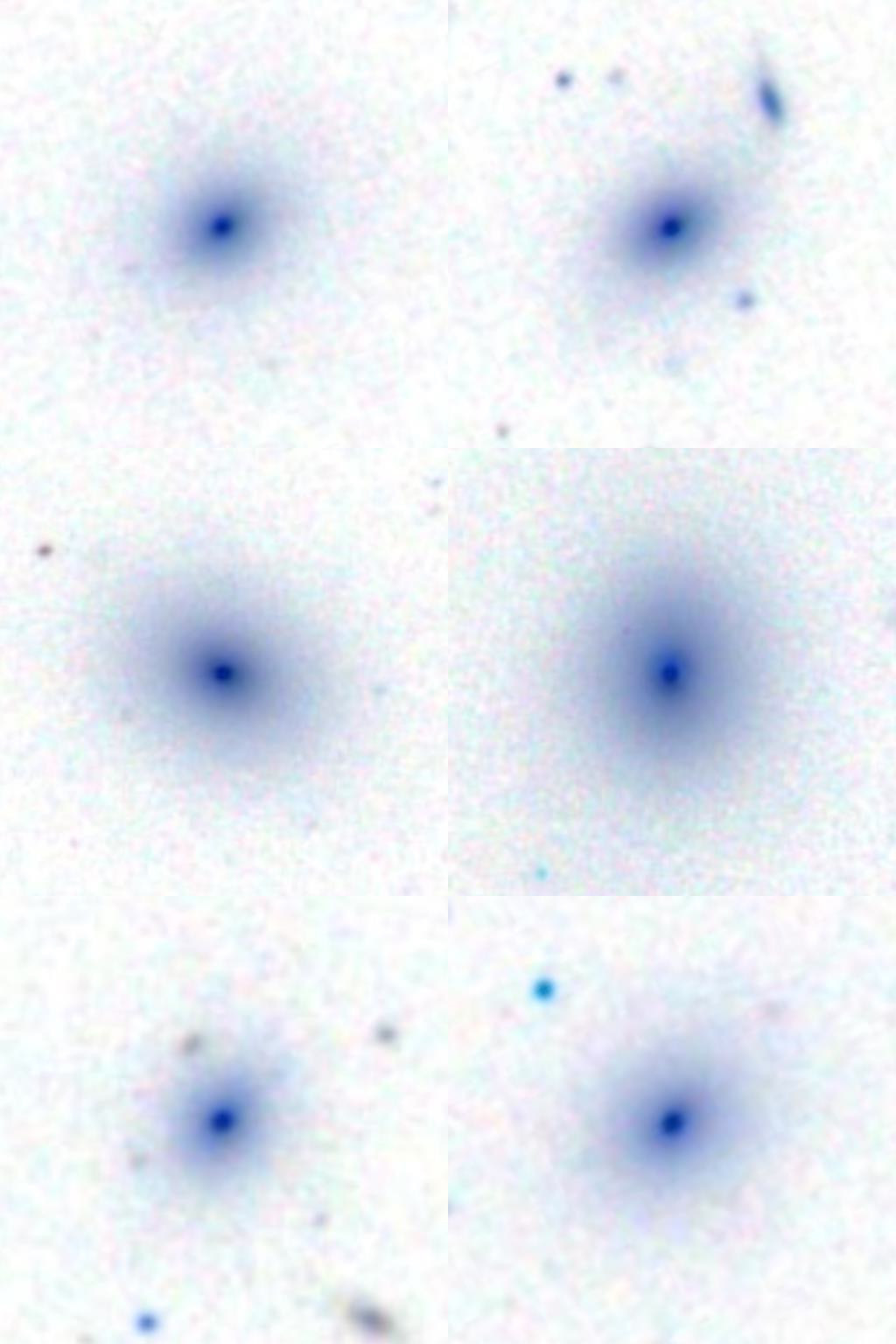}
		\includegraphics[width=.44\textwidth,valign=t]{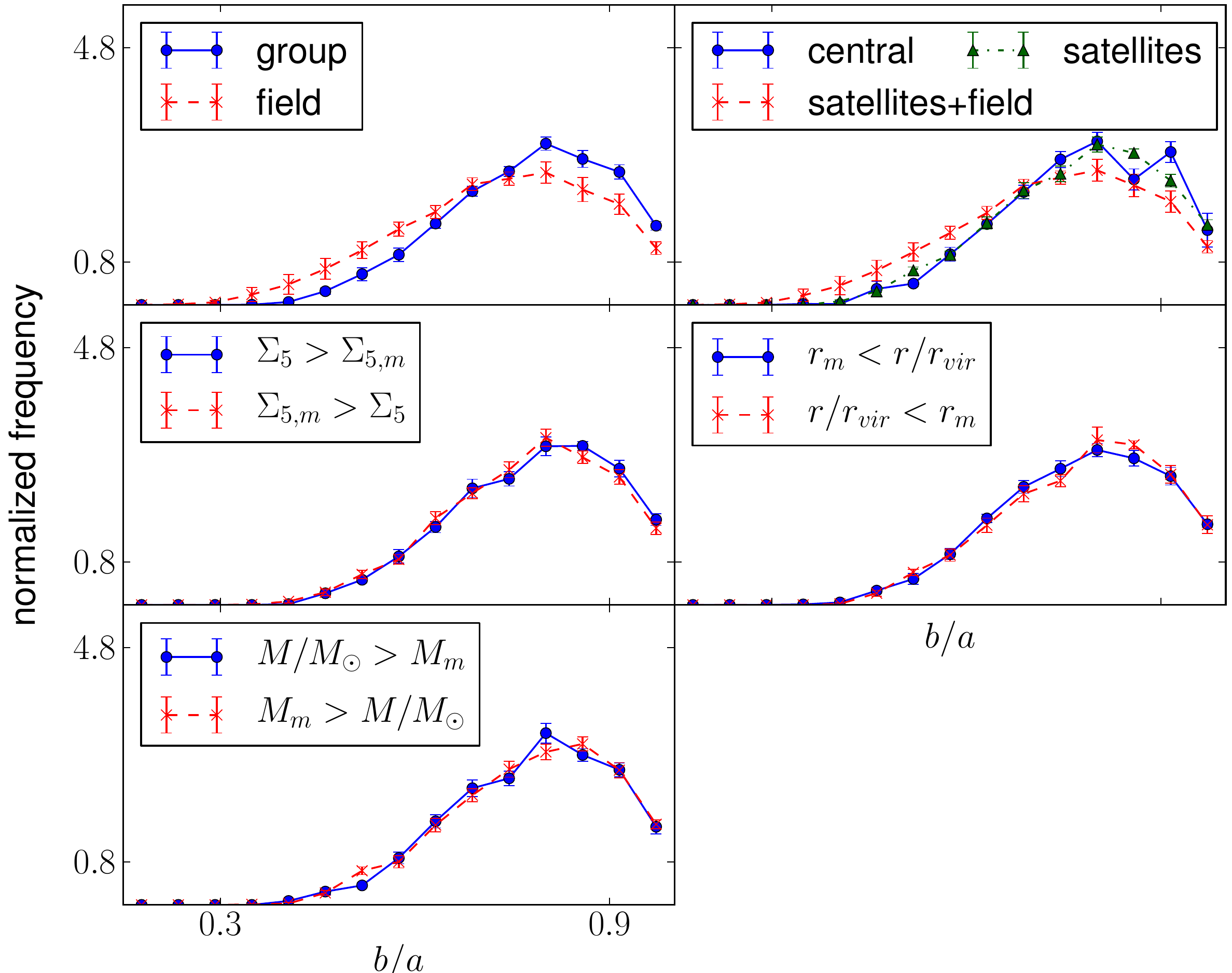}
		\includegraphics[width=.215\textwidth,valign=t]{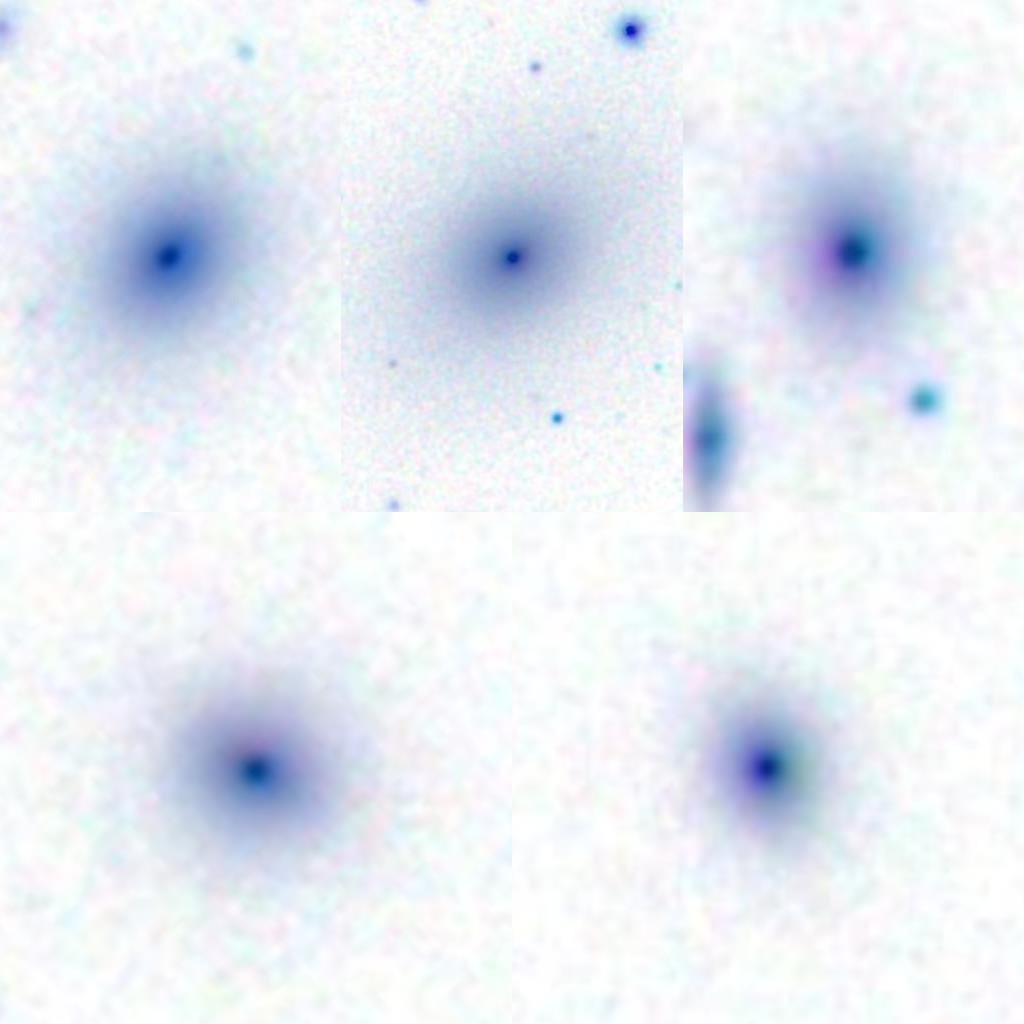}
		\caption{$b/a$ distributions for sub-samples divided by environmental properties along with images of sample galaxies from each sub-sample.
		The upper set of figures show the results for spirals, and the lower set for ellipticals.
		In both sets the top left panel shows samples of group and field galaxies separately. The top right panel shows the samples of central, satellite and satellite+field galaxies. The middle left panel shows the samples according to $\Sigma_5$. The middle right panel shows the samples divided according to $r/r_{vir}$ values. The bottom left shows samples divided by group mass. \label{fig:badist}}
   \end{center}
\end{figure*}

For the galaxies separated using discrete parameters (group or field, satellite or central) we assign galaxies to each sample so that the medians of the intrinsic properties used by RP13 (magnitude, colour an physical size) are similar between the samples that we are comparing. That is, the medians in intrinsic properties are similar between spiral group and spiral field galaxies, between centrals, satellites and satellite+field spirals. And the same for ellipticals. Table \ref{table:med_cut} shows the medians of intrinsic characteristics of each sample. Fig. \ref{fig:badist} shows the resulting distributions of $b/a$ for the selected sub-samples, separated between spirals and ellipticals using their Galaxy ZOO morphologies.

For galaxies selected by continuous parameters we take a different approach than for the other samples (where we cut the samples until their intrinsic properties are similar). We take the group sample without cuts, we separate in bins of Magnitude, colour and size, and for each bin we measure the median value of the parameter. We then select for each bin the galaxies above and below the median value of the environmental property; the median value is referred to as $\Sigma_{5,m}$ for projected density, which is a function of $M_r$, $g-r$ and $R_{50}$. For virial mass we use $M_m$, and $r_m$ for distance to the group centre in units of the virial radius. With this method we obtain samples with similar distributions of Magnitude, colour and size without excluding any galaxy.

The medians in magnitude, colour and size of these samples are also listed in Table \ref{table:med_cut} and Fig \ref{fig:badist} shows the $b/a$ distributions for these samples.

Fig \ref{fig:badist} also displays examples of galaxy images that belong to each sample. These galaxies where selected to have $b/a$ values close to the mean of the respective sub-sample.

\section{Results}
\label{results}

\begin{table*}
  \begin{center}
    \caption{Mean values of $\gamma$ and $\epsilon$ for all the analysed samples. Also are listed the $E_0$ values for the spiral samples.\label{table:results}}
	\begin{tabular}{lccccr}
	    & \multicolumn{3}{c}{Results for spirals}          & \multicolumn{2}{c}{Results for ellipticals}      \\
	  Sample&$E_0$ &$\langle\gamma\rangle$&$\langle\epsilon\rangle$&$\langle\gamma\rangle$&\multicolumn{1}{c}{$\langle\epsilon\rangle$}\\\hline
	  group                   &$0.218^{+0.068}_{-0.073}$&$0.352\pm0.017$&$0.126\pm0.01$ &$0.63\pm0.032$ &$0.097\pm0.017$\\
	  field                   &$0.159^{+0.048}_{-0.051}$&$0.349\pm0.011$&$0.099\pm0.018$&$0.51\pm0.033$ &$0.113\pm0.015$\\\hline
	  central                 &$0.11^{+0.168}_{-0.195}$ &$0.239\pm0.013$&$0.118\pm0.025$&$0.667\pm0.067$&$0.12\pm0.008$ \\
	  satellites              &$0.111^{+0.053}_{-0.073}$&$0.382\pm0.035$&$0.12\pm0.0107$&$0.61\pm0.026$ &$0.11\pm0.007$ \\
	  satellites+field        &$0.182^{+0.048}_{-0.052}$&$0.331\pm0.015$&$0.113\pm0.009$&$0.518\pm0.022$&$0.096\pm0.01$ \\\hline
	  $\Sigma_5>\Sigma_{5,m}$ &$0.195^{+0.055}_{-0.102}$&$0.269\pm0.029$&$0.146\pm0.046$&$0.647\pm0.027$&$0.113\pm0.009$\\
	  $\Sigma_{5,m}>\Sigma_5$ &$0.186^{+0.048}_{-0.055}$&$0.232\pm0.017$&$0.264\pm0.035$&$0.62\pm0.015$ &$0.099\pm0.017$\\\hline
	  $r_m>r/r_{vir}$         &$0.117^{+0.055}_{-0.083}$&$0.219\pm0.016$&$0.176\pm0.023$&$0.616\pm0.014$&$0.116\pm0.005$\\
	  $r/r_{vir}>r_m$         &$0.204^{+0.079}_{-0.072}$&$0.257\pm0.022$&$0.251\pm0.03$ &$0.611\pm0.02$ &$0.116\pm0.009$\\\hline
	  $M/M_{\sun}>M_m$       &$0.149^{+0.061}_{-0.07}$ &$0.249\pm0.022$&$0.186\pm0.02$ &$0.615\pm0.027$&$0.127\pm0.018$\\	
	  $M_m>M/M_{\sun}$	      &$0.144^{+0.06}_{-0.078}$ &$0.256\pm0.017$&$0.202\pm0.026$&$0.614\pm0.017$&$0.111\pm0.011$\\\hline
	\end{tabular}
  \end{center}
\end{table*}

\begin{figure*}
	\begin{center}
		\includegraphics[width=.45\textwidth]{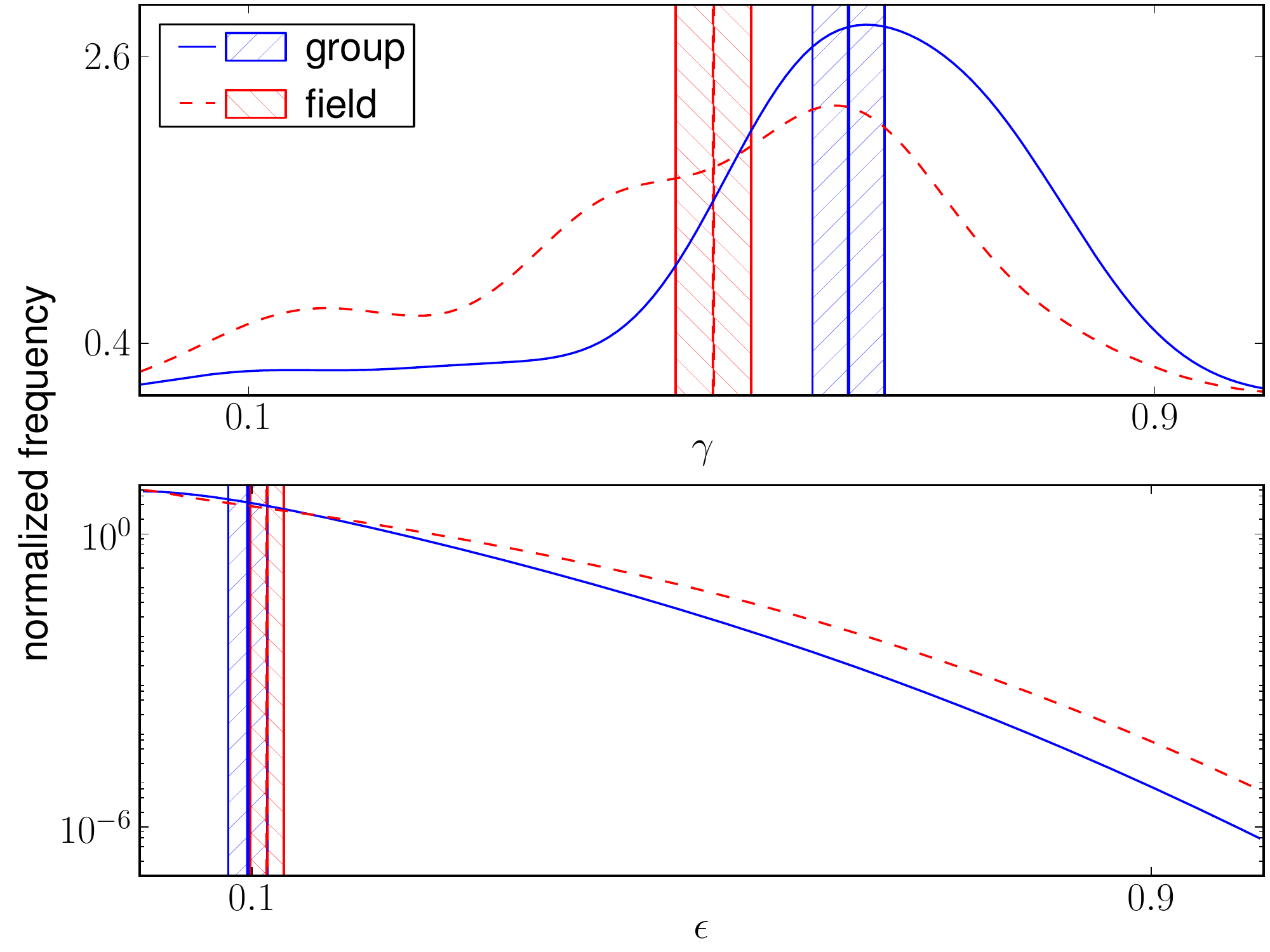}
		\includegraphics[width=.45\textwidth]{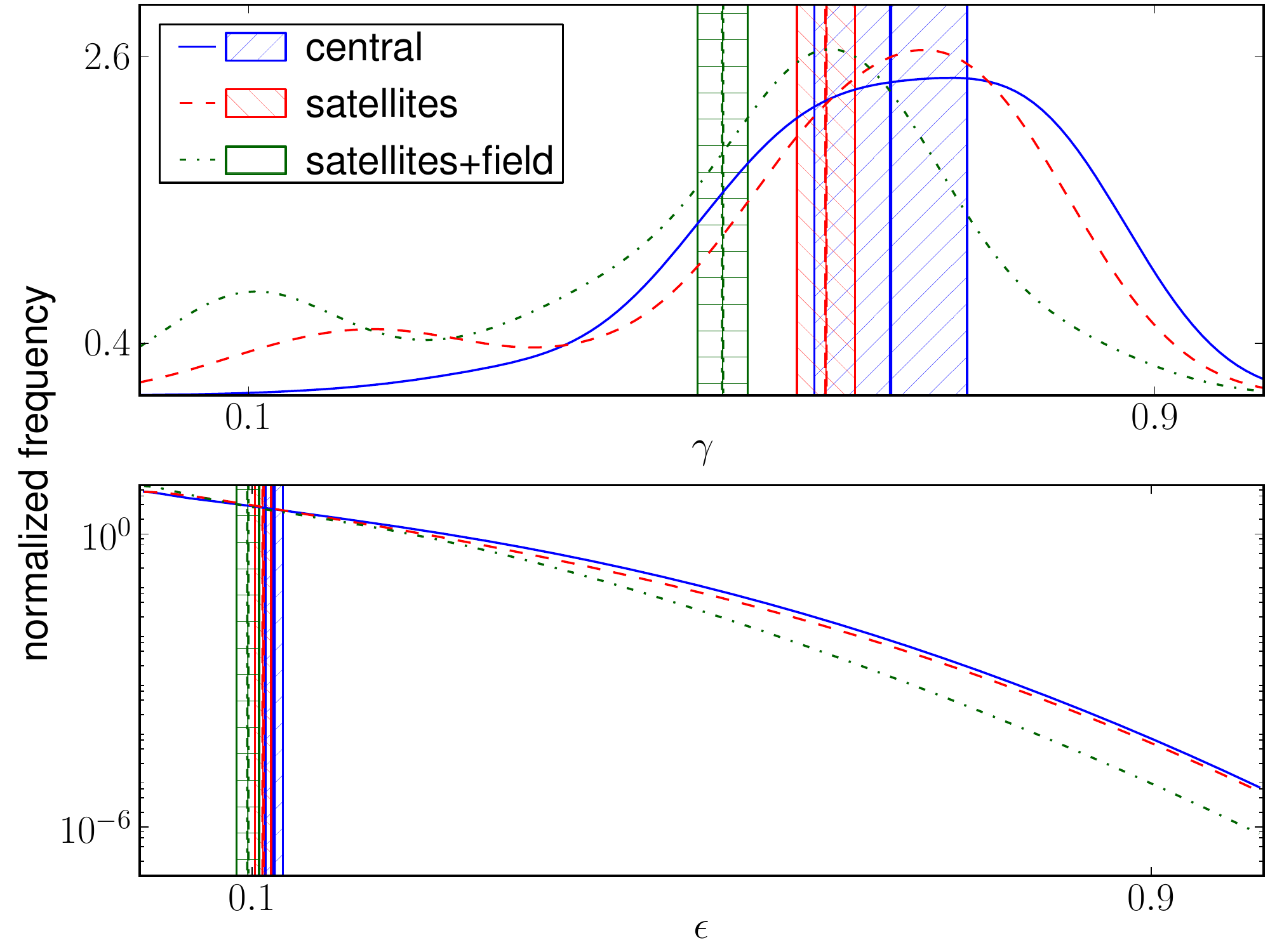}
		\caption{Distributions of $\gamma$ and $\epsilon$ for galaxy samples divided according to different properties. Top panels show $\gamma$ distributions and bottom panels show $\epsilon$ distributions. Left: group and field ellipticals. Right: central, satellite and satellite+field ellipticals. Vertical lines correspond to the mean values of the distributions and the dashed areas show the uncertainties of the mean values.\label{fig:exp_env}}
	\end{center}
\end{figure*} 

Fig. \ref{fig:exp_env} shows the distributions of $\gamma$ and $\epsilon$ for sub-samples of ellipticals. It can be seen in this figure important differences between the resulting distributions, in particular for ellipticals in groups and in the field. Table \ref{table:results} shows the mean values of $\gamma$ and $\epsilon$ obtained for all the samples, alongside with the values of $E_0$ for spirals. The top left panel in the right set of panels in Fig \ref{fig:badist} shows a clear difference between the group and field samples of ellipticals, this is confirmed by the results shown in the right panels in Fig \ref{fig:exp_env} in which we can see that the group galaxies tend to have a higher $\gamma$ than field ellipticals. That is, the ellipticals in groups are more spherical than field ellipticals.  In the case of spirals we do not find any relevant difference between the group and field galaxies. This can be confirmed by the data in Table \ref{table:results}, in which we can not see any important difference between the shape parameters or dust of those samples.

Fig \ref{fig:badist} shows that there is no important difference between satellite and central ellipticals. The right panels in Fig \ref{fig:exp_env} do not show any difference in the  $\gamma$ distributions of centrals and satellites. The satellite+field ellipticals shows a difference in $\gamma$ with the other two samples, but this difference could be due to the difference between group and field ellipticals.

\begin{figure*}
	\begin{center}
		\includegraphics[width=.45\textwidth]{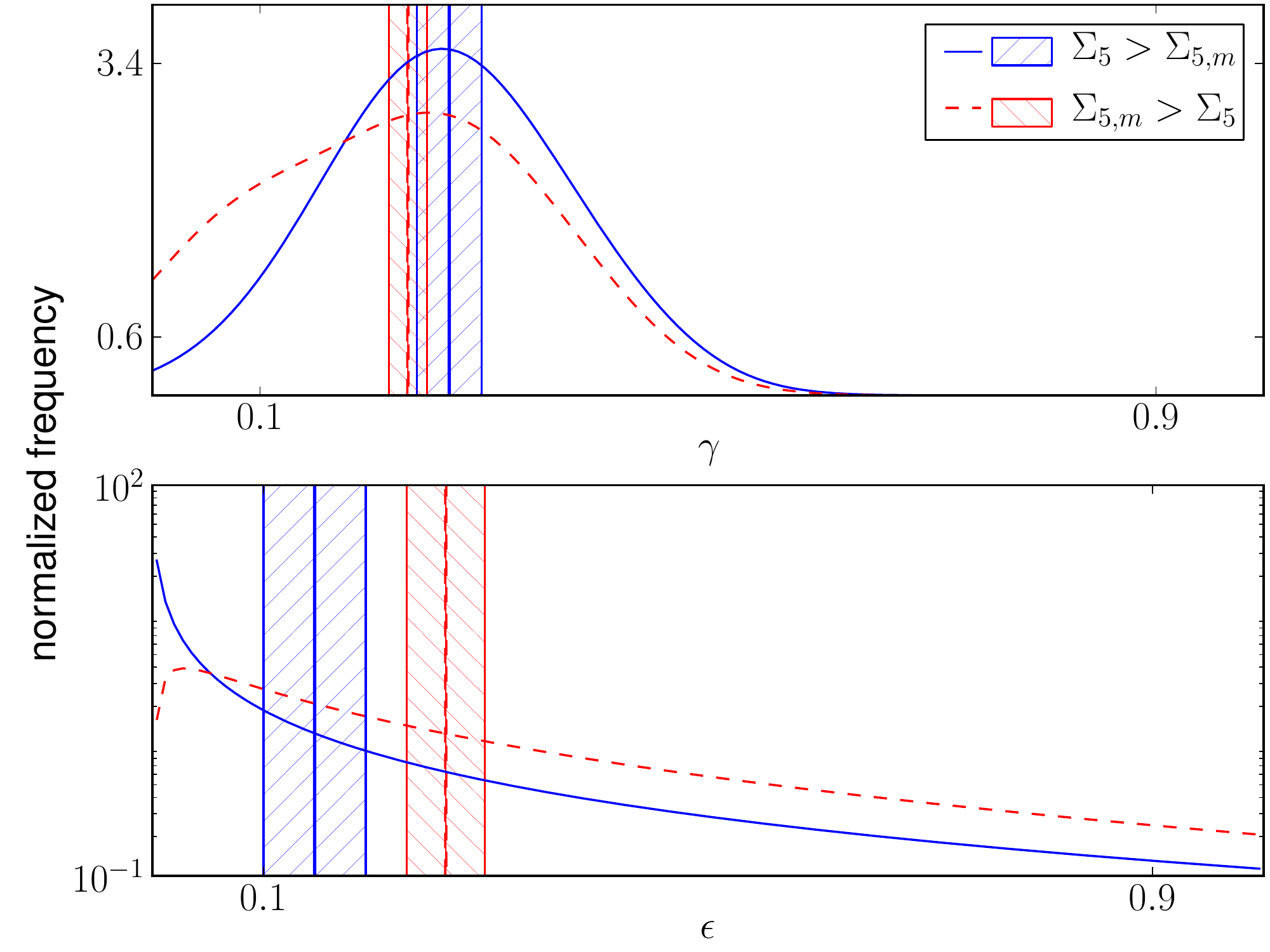}
		\includegraphics[width=.45\textwidth]{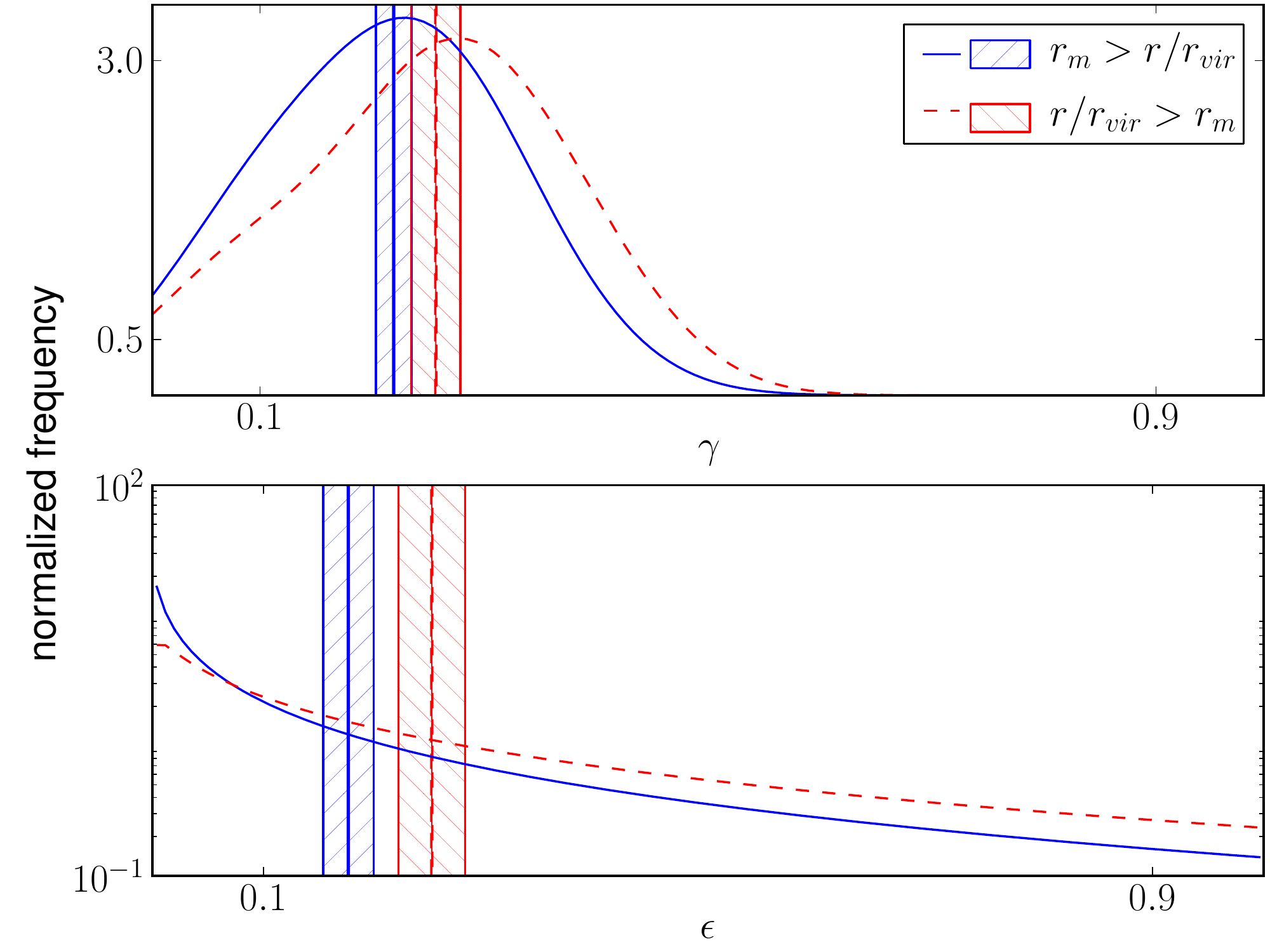}
		\caption{Distributions of $\gamma$ and $\epsilon$ for samples separated by continuous parameters. Top panels show the $\gamma$ distributions and bottom panels show the $\epsilon$ distributions. Left: spirals separated by $\Sigma_5$. Right: spirals separated $r/r_{vir}$. \label{fig:ge_sigma}}
	\end{center}
\end{figure*}

Fig \ref{fig:ge_sigma} shows the fit best parameters for samples selected by continuous properties that show significant differences between the distributions of $\gamma$ or $\epsilon$. Fig. \ref{fig:badist} shows that there is no important difference between the galaxies with $\Sigma_5>\Sigma_{5,m}$ and $\Sigma_{5,m}>\Sigma_5$ for ellipticals, and shows just a small difference for spirals. For these spirals there is an important difference in the $\epsilon$ distribution, in which the galaxies in denser environments tend to have a rounder disk than galaxies in less dense environments. This difference is consistent with the results shown in Fig. \ref{fig:exp_env} in which spiral galaxies closer to the centre of their host group tend to have a smaller value of $\epsilon$ than spirals located further from centre. For ellipticals we do not find any important difference between galaxies in environments with different density (see Table \ref{table:results}).

The middle right panel in both sets of panels of Fig. \ref{fig:badist} shows that there is no significant difference between galaxies located relatively near to the centre of mass of the group for spirals, and galaxies that are far from the centre, in terms of the virial radius. Fig. \ref{fig:exp_env} shows that spirals do not show a difference in $\gamma$, but there is a difference in $\epsilon$, where the spirals that are near to the centre of the group tend to have a smaller value of $\epsilon$ that the other spirals in groups. These results shows that spirals that are relatively near to the centre tend to have a more circular disc. This is consistent with the results using $\Sigma_5$ and the central-satellites samples. For ellipticals we do not find any relation between shape and the distance to the centre of mass of the group.

In the case of samples separated by virial mass of the group, the shapes of the galaxies do not show any difference between the high and the low mass group samples for spirals or ellipticals. Table \ref{table:results} show that there is no difference between the shape parameters or the dust extinction of any of these samples.

\subsection{Central and satellite spirals}
\label{sub:central}

\begin{figure}
	\begin{center}
		\includegraphics[width=.45\textwidth]{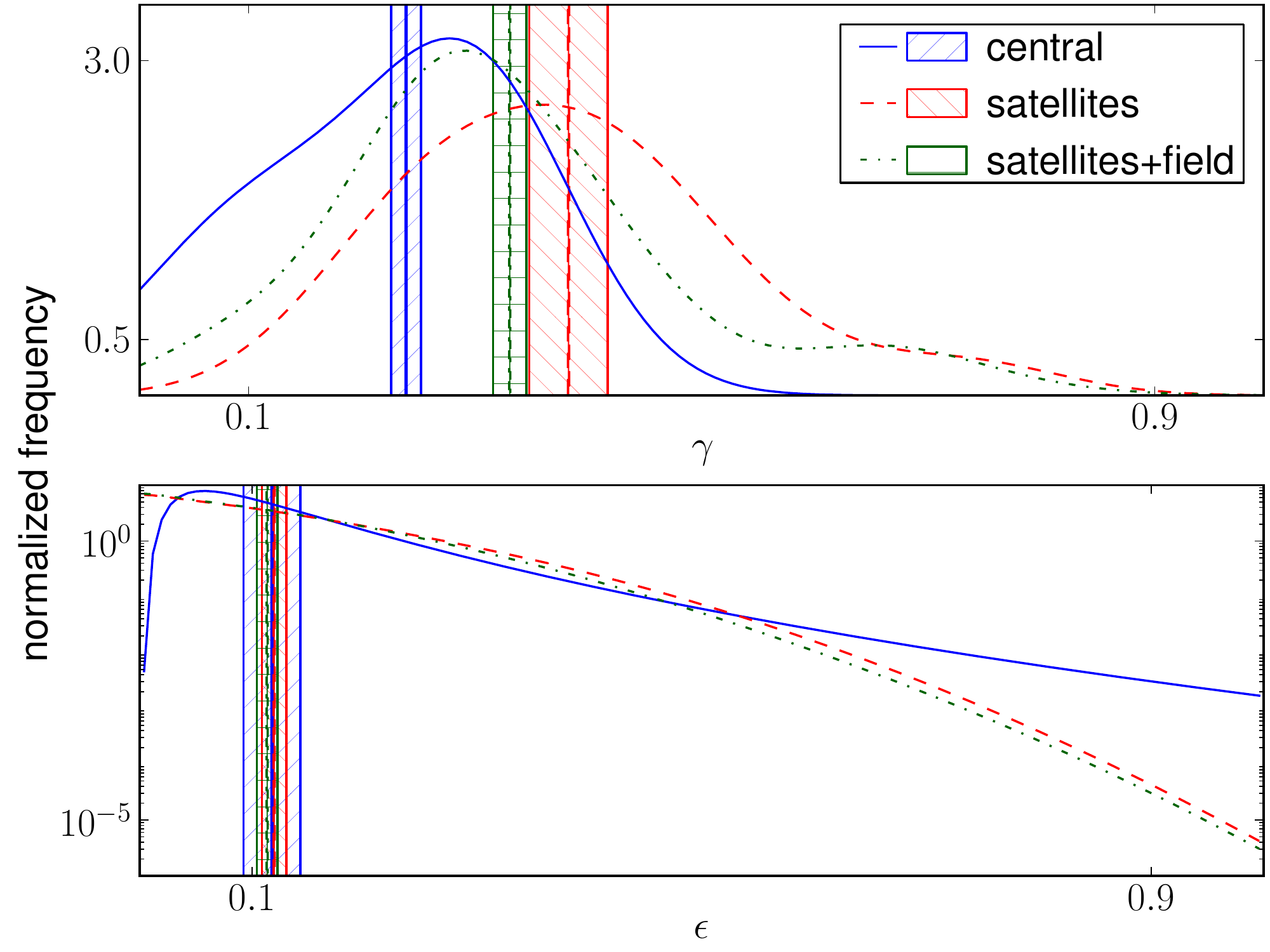}
		\includegraphics[width=.45\textwidth]{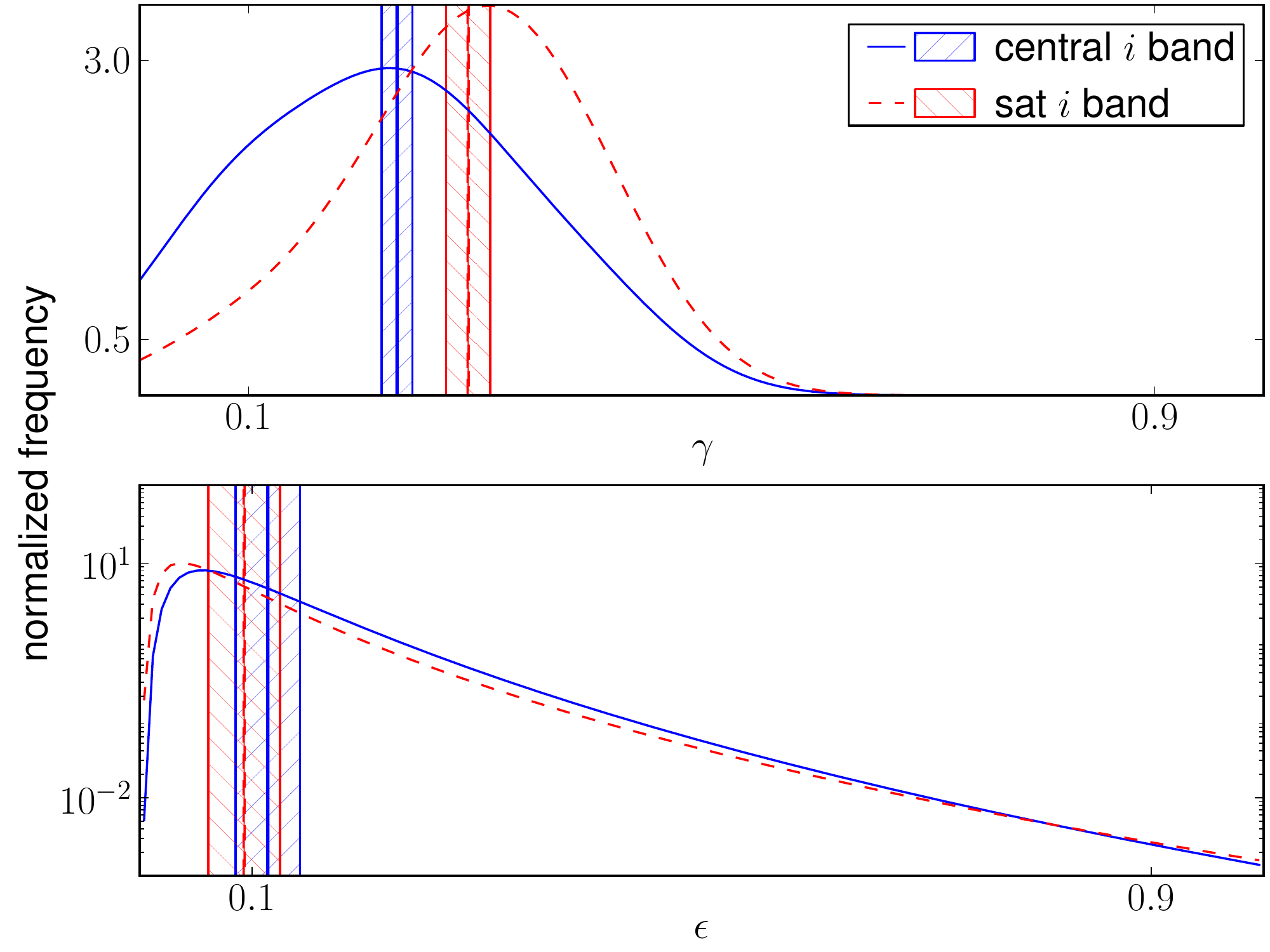}
		\caption{Distribution of $\gamma$ and $\epsilon$ for spirals separated into central, satellites and satellites+field samples. The upper set of panels contains the distributions calculated using the $r$ band data. The lower set of panels contains the distributions that we obtain using the $i$ band data (do not include the satellite+field sample).\label{fig:central}}
	\end{center}
\end{figure}

Among the 3D shape distributions obtained, we argue that those of central and satellite spirals deserve a further analysis.   Fig \ref{fig:central} shows the $\gamma$ and $\epsilon$ distributions for these samples, which exhibit no important differences. However, $\gamma$ distributions suggest that central galaxies tend to be thinner than central spirals. This trend is also present in the satellites+field sample. Summary of these results are also listed in Table \ref{table:results}.

In order to further test these results, we have performed a similar analysis using the $i$ band (both for the $b/a$ parameter and galaxy magnitudes) which is less affected by seeing \citep{dr1}. The results are also given in Fig \ref{fig:central}  showing the same trend that those in the $r$-band distribution.

Since it is expected that events affecting morphology were more frequent and violent in central spirals than in satellites, the shape of central galaxies would, in principle, tend to be more spherical than those of satellites. However, \cite{alatalo}, using galaxies from the ATLAS$^{\textmd{3D}}$ survey \citep{atlas3d}, have recently found that the ratio between $^{12}$CO(1-0)/$^{13}$CO(1-0) decrease with the density of the environment \citep[see also][]{crocker}. In these works it is proposed that the gas is confined to a thinner disc due to the pressure from the intra cluster/group medium so that star formation would also occur in a thinner disc. This fact could explain our observations of central spirals thinner than satellites.

\begin{figure*}
	\begin{center}
		\includegraphics[width=.325\textwidth]{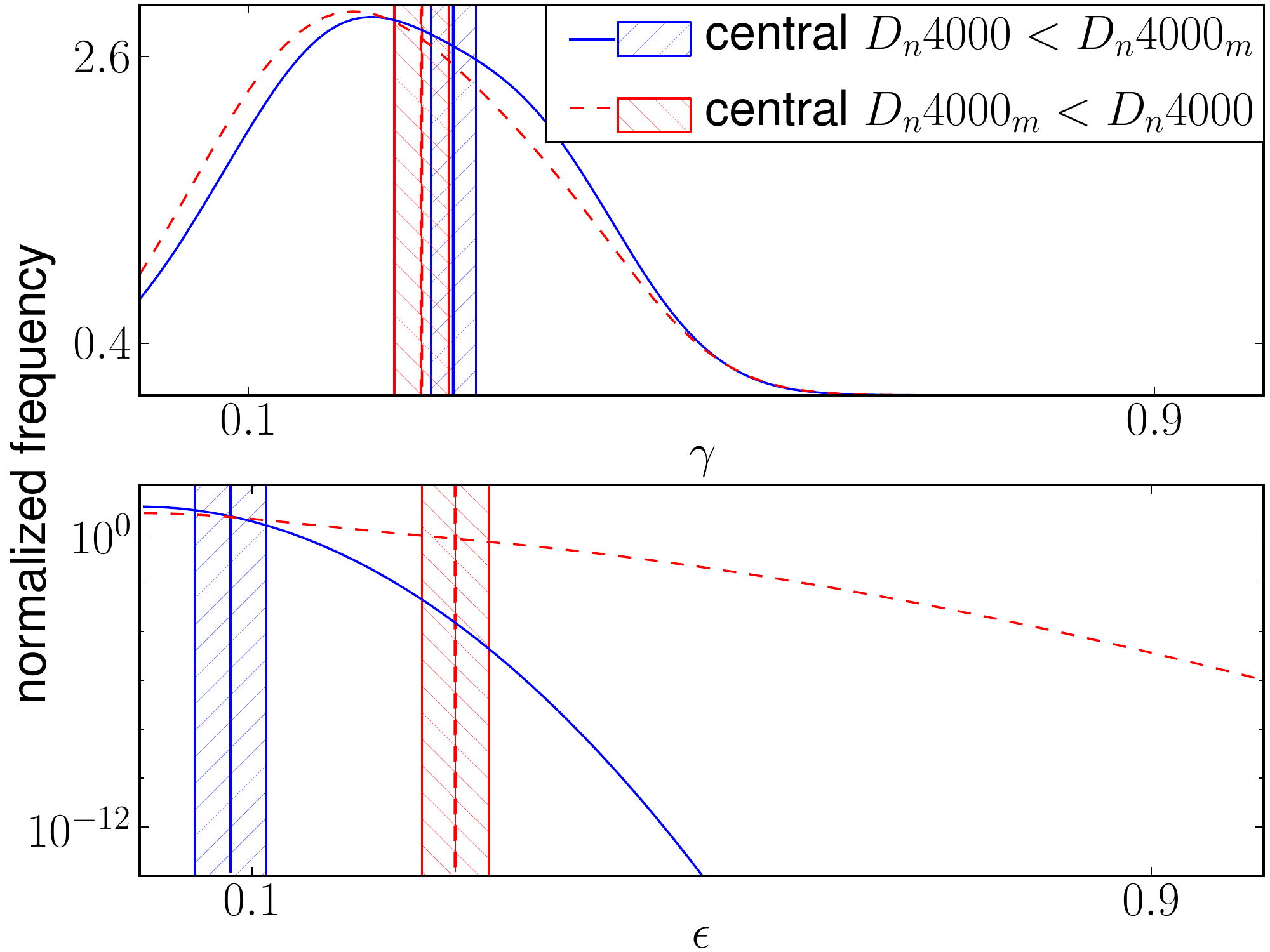}
		\includegraphics[width=.325\textwidth]{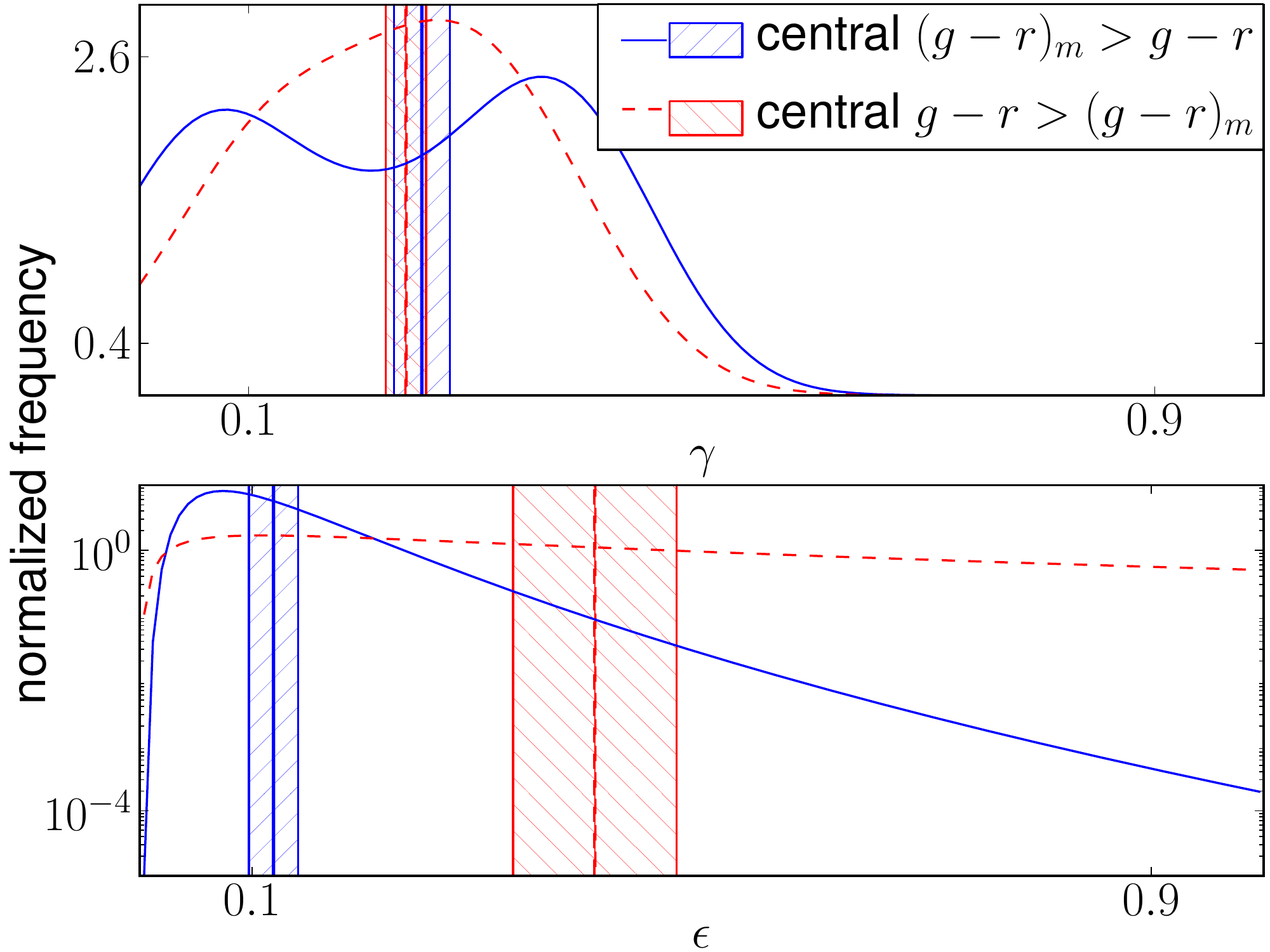}
		\includegraphics[width=.325\textwidth]{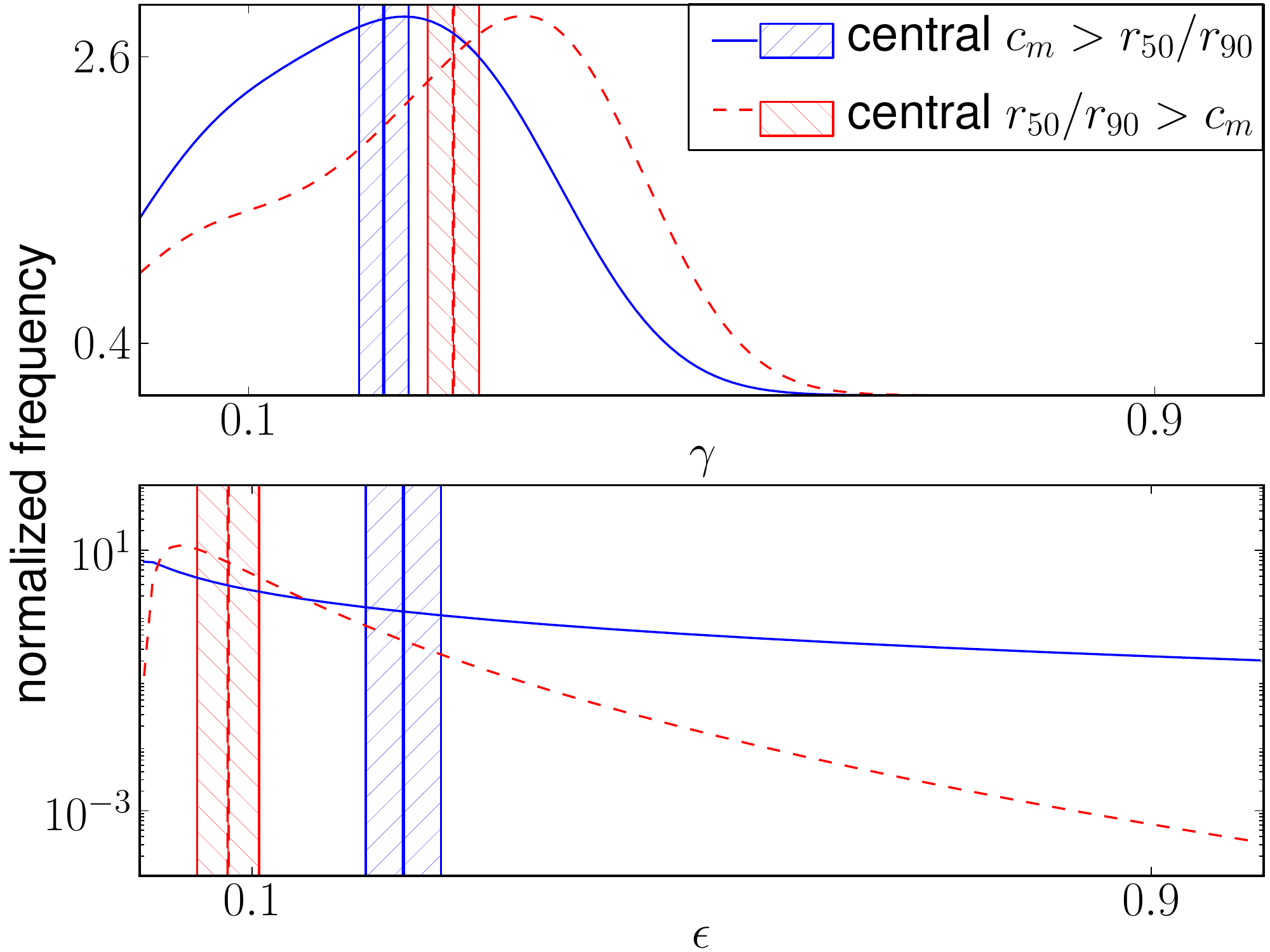}
		\includegraphics[width=.325\textwidth]{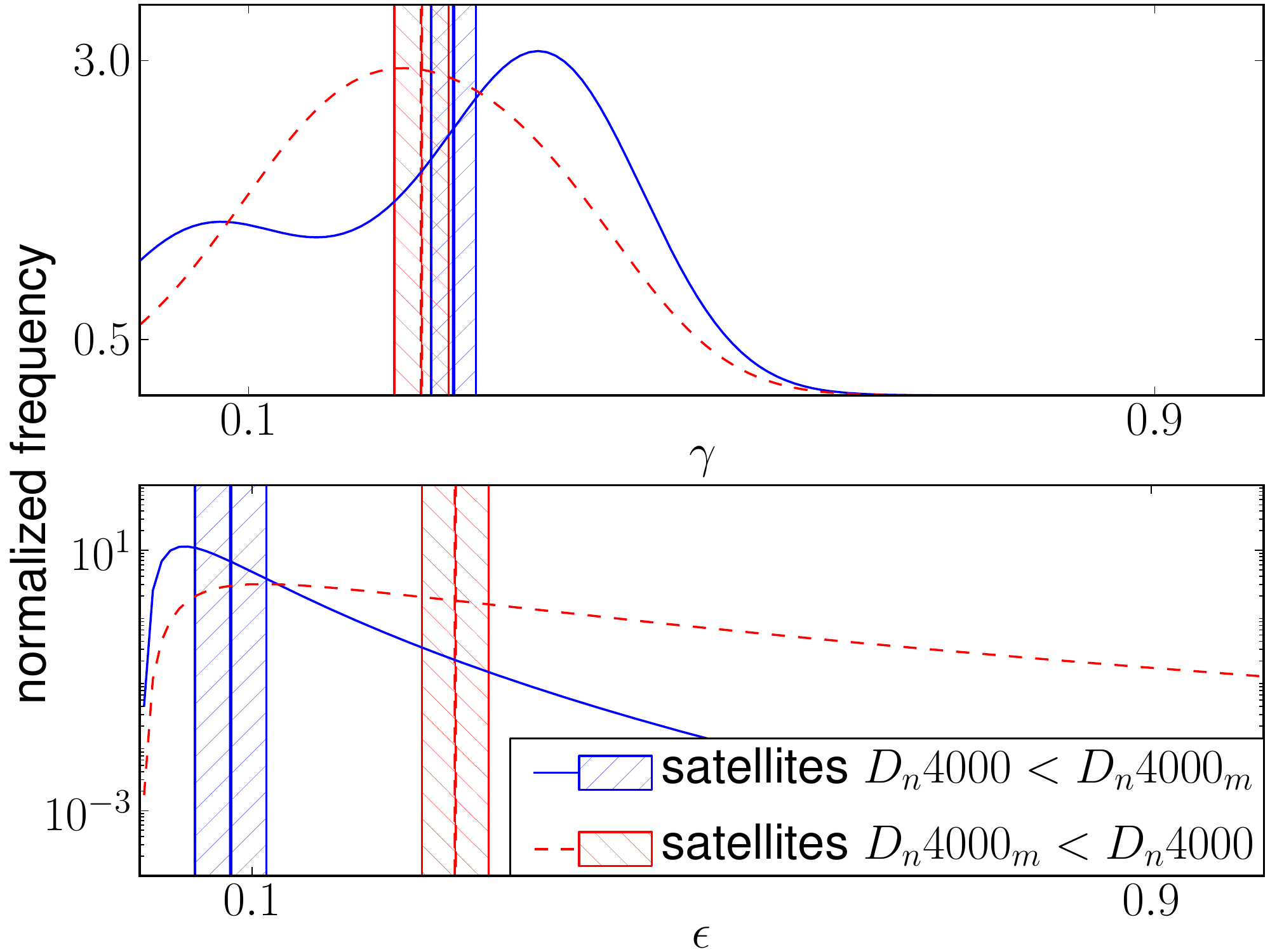}
		\includegraphics[width=.325\textwidth]{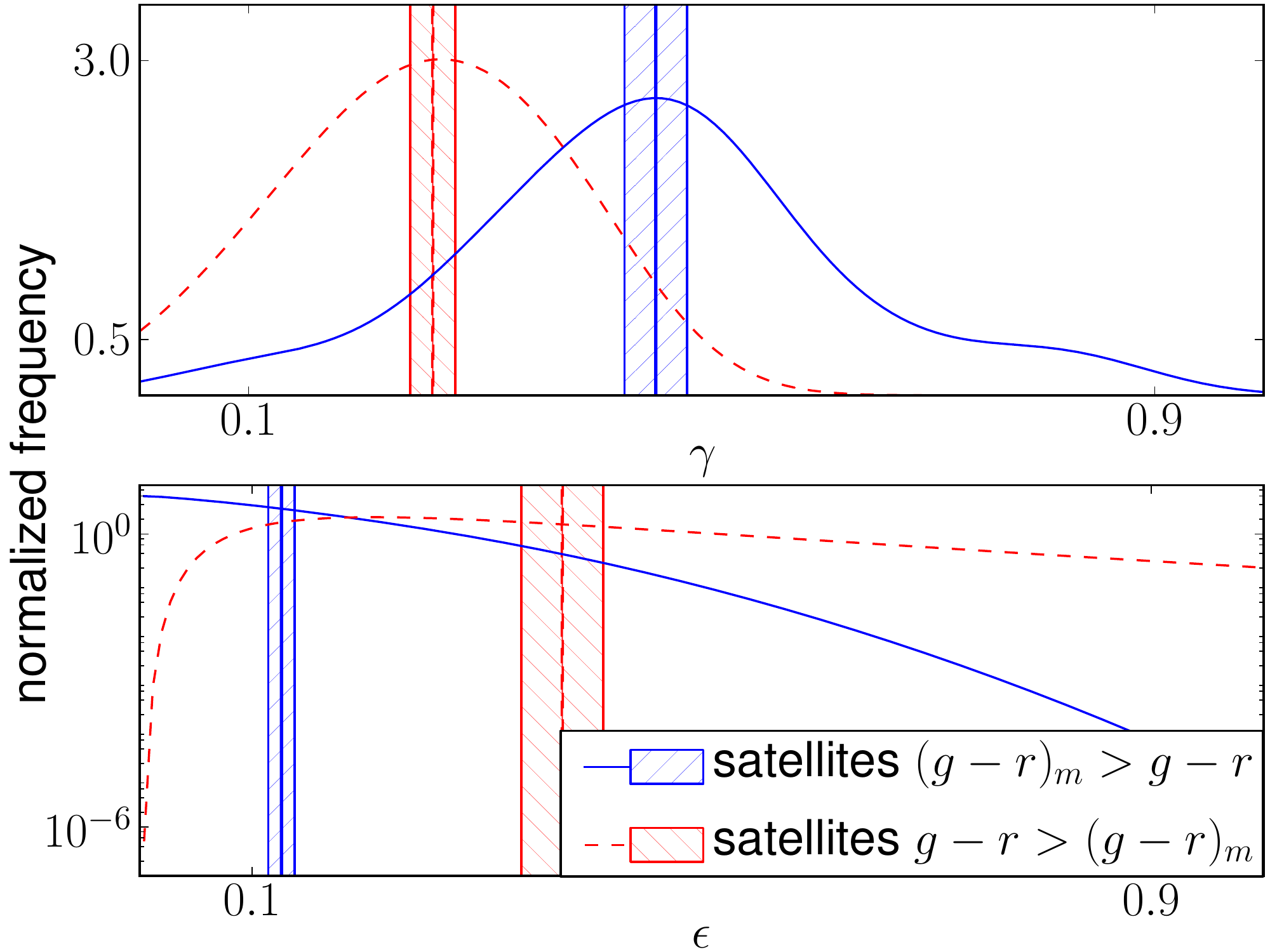}
		\includegraphics[width=.325\textwidth]{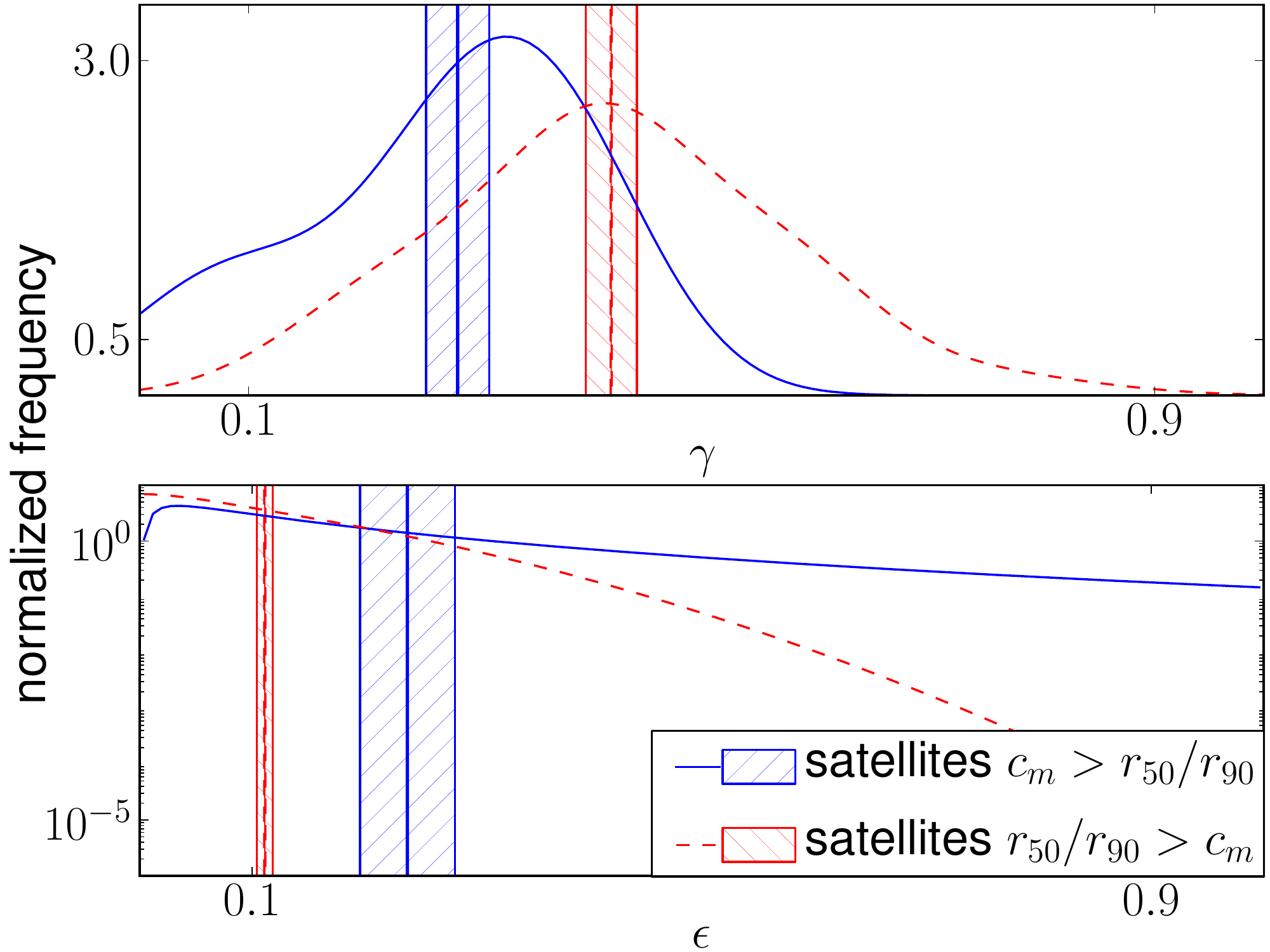}
		\caption{$\gamma$ and $\epsilon$ distributions for the samples of central and spirals separated by different intrinsic properties. The upper panels shows the distributions of central galaxies. the lower panels show the distribution for satellites galaxies. LEFT: Samples separated by $D_n(4000)$. MIDDLE: Galaxies separated by $g-r$ colour. RIGHT: Galaxies separated by concentration.\label{fig:centralthings}}
	\end{center}
\end{figure*}

In order to further test this hypothesis, we sub-split the central and satellite spiral samples using parameters related to the stellar age,  star formation and the structure of the galaxy ($D_n4000$, the strength of the break at 4000 $\mathring{\textmd{A}}$ see, $g-r$ colour and concentration in the $r$ band). The $D_n4000$ and concentration sub-samples were built following the same method that those applied to continuous variables (such as $\Sigma_5$) and the $g-r$ case was obtained splitting in equal number sub-samples . Fig \ref{fig:centralthings} shows the results obtained and Table \ref{table:extra} list the values of $E_0$, $\langle \gamma \rangle$ and $\langle \epsilon \rangle$.

The results for $D_n4000$ (Left panels in Fig \ref{fig:centralthings}) show no important difference between the $\gamma$ distributions, but galaxies with a lower $D_n4000$ values tend to have a lower $\epsilon$. This implies that for both central and satellite, galaxies with a younger stellar populations tend to have a more circular disc.

The distributions for the samples divided according $g-r$ (middle panels if Fig \ref{fig:centralthings}) show that, for central galaxies, the bluer ones tend to have a more circular disc, for satellites, the redder galaxies show a thinner and less circular disc. In the case of concentration (right panels in Fig \ref{fig:centralthings}) in both cases the less concentrated galaxies tend to have a thinner and less circular disc.

The colour results are in good agreement with the hypothesis of \citeauthor{crocker} and \citeauthor{alatalo}, since the redder  galaxies are those expected to have suffered stronger confinement by the warm/hot intra-group medium, are those with thinner discs. The redder, older ones are expected to have formed and evolved under this confinement, while the blue ones are expected to have star formation activity post-mergers which possibly heated the disc causing higher $\gamma$ values. We stress the fact that colour may better represent the global age of the galaxy disc since spectroscopic derived parameters contain information of the central bulge of this relatively near-by SDSS galaxy sub-samples. It is important to notice that the $\epsilon$ results, where the bluer galaxies have a rounder disc, has also been already observed in RP13, where the environment has not been taken into account, suggesting that the difference in $\epsilon$ values is likely due to intrinsic  galaxy properties. Similarly, RP13 found no important differences in $\gamma$ distributions using colour cuts.

The concentration results, on the other hand, provide suitable predictions, where disc dominated galaxies (those with lower concentration parameters) tend to be thinner than bulge dominated spirals, for both centrals and satellites. We argue that spirals with larger bulges are likely to have thicker discs given their possible common origin via mergers of substructures \citep{quinn,aguerri}.

\begin{table*}
	\begin{center}
		\caption{$E_0$ values, $\langle \gamma \rangle$ and $\langle \epsilon \rangle$ values for the sub samples created from the central and satellite sample.\label{table:extra}}
		\begin{tabular}{lccr}
			sample & $E_0$ & $\langle \gamma \rangle$ & \multicolumn{1}{c}{$\langle \epsilon \rangle$} \\\hline
			central $i$ band &  $0.298^{+0.115}_{-0.13}$ & $0.231\pm0.014$ & $0.114\pm0.029$ \\
			satellites $i$ band & $0\pm0$ & $0.294\pm0.02$ & $0.093\pm0.032$ \\\hline
			central $D_n4000<D_n4000_m$ & $0.176^{+0.151}_{-0.148}$ & $0.214\pm0.038$ & $0.129\pm0.039$ \\ 
			central $D_n4000_m<D_n4000$ & $0.746^{+0.134}_{-0.137}$ & $0.273\pm0.045$ & $0.24\pm0.033$ \\
			satellites $D_n4000<D_n4000_m$ & $0\pm0$ & $0.281\pm0.02$ & $0.081\pm0.032$ \\
			satellites $D_n4000_m<D_n4000$ & $0.418^{+0.057}_{-0.066}$ & $0.253\pm0.024$ & $0.281\pm0.03$ \\\hline
			central $g-r_m>g-r$ & $0.368^{+0.145}_{-0.149}$ & $0.253\pm0.025$ & $0.119\pm0.022$ \\
			central $g-r>g-r_m$ & $0.833^{+0.156}_{-0.131}$ & $0.239\pm0.018$ & $0.405\pm0.073$ \\
			satellites $g-r_m>g-r$ & $0.201^{+0.116}_{-0.124}$ & $0.46\pm0.028$ & $6 0.126\pm0.012$ \\
			satellites $g-r>g-r_m$ & $0.293^{+0.063}_{-0.133}$ & $0.263\pm0.02$ & $0.376\pm0.036$ \\\hline
			central $c_m>r_{50}/r_{90}$ & $0.357^{+0.123}_{-0.11}$ & $0.219\pm0.022$ & $0.235\pm0.034$ \\
			central $r_{50}/r_{90}>c_m$ & $0.347^{+0.177}_{-0.118}$ & $0.281\pm0.023$ & $0.079\pm0.027$ \\
			satellites $c_m>r_{50}/r_{90}$ & $0.217^{+0.11}_{-0.141}$ & $0.285\pm0.028$ & $0.238\pm0.042$ \\
			satellites $r_{50}/r_{90}>c_m$ & $0.18^{+0.067}_{-0.117}$ & $0.42\pm0.023$ & $0.111\pm0.007$ \\\hline
		\end{tabular}
	\end{center}
\end{table*}

\section{Conclusions}
\label{conclusions}

We have extended the analysis of the intrinsic shape of galaxies in RP13 to samples selected by environmental properties (galaxies in groups and field galaxies, central and satellites, near and far from the centre of mass of the group, in different density environments and in groups with different mass) to determine if the environment alone has a role to play in the distribution of galaxy shapes.

We find that spiral galaxies in groups are not different in shape than field spirals. However, central spirals contain thinner disc than satellite spirals, and galaxies relatively closer to the centre of mass of the group tend to have a rounder disc than spirals that are farther away. We also found that spirals in environments with a higher projected density also tend to have a rounder disk than galaxies located low density regions.

For elliptical galaxies, we find that group ellipticals are more spherical than field ones, and the same for central ellipticals compared to satellite ellipticals. For samples of ellipticals selected according to the density of the environment, the distance to the centre and the mass of the group, we do not find  any important difference.

In conclusion, we find that for ellipticals, their shape in cluster and field environments are different, but when the ellipticals are restricted to groups variations of the environment does not play a major role on their shape. For spirals, on the other hand, the environment has an important influence on their shape,  specifically the position of the galaxy within the cluster. The mass of the group itself has no apparent influence.

For ellipticals, the result of more spherical galaxies in groups rather than the field could be explained by the fact that elliptical galaxies in groups are exposed to more events that affect their shapes (mergers, gravitational interactions, etc). These events affect the galaxies mostly isotropically, and lead to more spherical galaxies.
Perhaps more difficult to understand is the fact that spiral galaxies in denser environment environments tend to show a
thinner and rounder disc. Nevertheless \cite{crocker}, studied the line ratios of molecular gas in galaxies from the ATLAS$^{\textmd{3D}}$ survey \citep{atlas3d}, and propose that galaxies in environments with higher density tend to suffer high pressure due to the hot intra cluster/group medium, which leads to a gas confined in a thinner disc. This would explain the apparent (but very weak) decreasing relation they found between the ratio of $^{12}$CO(1-0)/$^{13}$CO(1-0) and the density.  This relation was also seen by \citep{alatalo}, and may cause star formation to also occur in a thinner disc.

\section*{Acknowledgements}

We want to thanks Chris Power, for his comments who help us to focus on the important results, and to give the text a more clear development.

This work was supported by Consejo Nacional de Investigaciones Cient\'ificas y T\'ecnicas (CONICET). NP was supported by Proyecto Fondecyt Regular 1150300.

The Geryon cluster housed at the Centro de Astro-Ingenieria UC was used for the calculations performed in this paper. The BASAL PFB-06 CATA, Anillo ACT-86, FONDEQUIP AIC-57, and QUIMAL 130008 provided funding for several improvements to the Geryon cluster.

Funding for SDSS-III has been provided by the Alfred P. Sloan Foundation, the Participating Institutions, the National Science Foundation and the US Department of Energy Office of Science. The SDSS-III website is \url{http://www.sdss3.org/}.

SDSS-III is managed by the Astrophysical Research Consortium for the Participating Institutions of the SDSS-III Collaboration including the University of Arizona, the Brazilian Participation Group, Brookhaven National Laboratory, University of Cambridge, Carnegie Mellon University, University of Florida, the French Participation Group, the German Participation Group, Harvard University, the Instituto de Astrofisica de Canarias, the Michigan State/Notre Dame/JINA Participation Group, Johns Hopkins University, Lawrence Berkeley National Laboratory, Max Planck Institute for Astrophysics, Max Planck Institute for Extraterrestrial Physics, New Mexico State University, New York University, Ohio State University, Pennsylvania State University, University of Portsmouth, Princeton University, the Spanish Participation Group, University of Tokyo, University of Utah, Vanderbilt University, University of Virginia, University of Washington and Yale University.

\bibliographystyle{mnras}
\footnotesize
\bibliography{references}

\label{lastpage}
\end{document}